\documentclass[conference]{IEEEtran}
\IEEEoverridecommandlockouts
\usepackage{amsmath} % 很多数学公式,  例如组合数等等,  都要用到这个宏.  注意：想用\eqref, 也要引用这个宏
\usepackage{amssymb} % 一些特殊的字体（如\mathbb等）需要这个宏.
\usepackage{MnSymbol}
\usepackage{extarrows}% 长等号上下标注,  长箭头上下标注
\usepackage{enumerate} % 用来调整enumerate环境,  至少想要修改枚举的编号需要使用这个宏
\usepackage{bm} % 加粗字体包
\usepackage{booktabs}
\usepackage{array}
\usepackage{graphicx} % 插入图片
\usepackage{subfigure} % 插入子图
\usepackage{multirow} % 表格跨行跨列

\usepackage{cite}
\usepackage{amsmath,amsfonts}
\usepackage{algorithmic}
\usepackage{array}
\usepackage[caption=false,font=normalsize,labelfont=sf,textfont=sf]{subfig}
\usepackage{textcomp}
\usepackage{stfloats}
\usepackage{url}
\usepackage{verbatim}
\usepackage{xcolor}
\usepackage{graphicx}
\newtheorem{theorem}{Theorem}[section]
\newtheorem{definition}{Definition}[section]

\newtheorem{lemma}{Lemma}[section]

\newtheorem{proposition}{Proposition}[section]

\newtheorem{example}{Example}[section]
\newtheorem{remark}{Remark}[section]
\newcommand{\pf}{\textbf{Proof: }}
\newcommand{\e}{\hfill$\blacksquare$}

\usepackage{epstopdf}

\allowdisplaybreaks[4] % 允许公式跨页,  [a]表示跨页力度,  a越大,  跨页力度越大

\date{}
\def\BibTeX{{\rm B\kern-.05em{\sc i\kern-.025em b}\kern-.08em
		T\kern-.1667em\lower.7ex\hbox{E}\kern-.125emX}}
\usepackage{balance}
\begin{document}

	\title{A Method to Reduce the Complexity of Computing
the Complete Weight Distribution of Polar Codes
		%	should not be used}
	}

	\author{\fontsize{10.3pt}{\baselineskip}\selectfont Zhichao Liu$^{*\dagger}$, Zhiming Ma$^{*\dagger}$ and Guiying Yan$^{*\dagger}$\\
		$^*$University of Chinese Academy and Sciences, Beijing, China\\
		$^\dagger$Academy of Mathematics and Systems Science, CAS, Beijing, China\\

		Email: liuzhichao20@mails.ucas.ac.cn, mazm@amt.ac.cn, yangy@amss.ac.cn
		\thanks{This work was supported by the National Key R\&D Program of China (No. 2023YFA1009602).}
	}
	
	\maketitle
	
	\begin{abstract}
		The code spectrum of polar codes is crucial to the performance of polar codes. Based on the lower-triangular affine group (LTA) of decreasing monomial codes and the one-variable descendance (ovd) relation, we define a new subgroup of LTA which can find more cosets with the same weight distribution. Using this algebraic structure, we further reduce the complexity by proofing the group action on a coset set is transitive. Our method is an enhanced version of previous research, and the complexity of most cases can be reduced exceeding several times. 
		
	\end{abstract}
	\section{Introduction}\label{section1}
	Polar codes, invented by Arıkan \cite{arikan2008}, have been proved to be capacity achieving under successive cancellation (SC) decoding. And when the code length $N$ tends to infinity, polarization will occur: one part of the channel will tend to be perfect, and the other part of the channel will tend to be pure-noise. In order to achieve the performance of maximum likelihood (ML) decoding, successive cancellation list (SCL) decoding is proposed in \cite{scl} \cite{list}, and it is equivalent to ML decoding when the list is very large.
	
	The code spectrum of linear codes is closely related to the ML performance of the codes, which can be estimated by union bound. However, most of the linear code spectrum calculations have been proved to be NP hard \cite{NP}, and only the code spectrum with few digits is known, such as Hamming codes and Golay codes. The polar code spectrum needs to be calculated to obtain the weight distribution of $2^{M(\mathcal{C})}$ cosets, where $M(\mathcal{C})=|\{i\in \mathcal{I}|i<\tau\}|$, $I$ is the information bits set and $\tau$ is the last frozen bit. The number of the cosets to be computed can be greatly reduced by the ovd relation of $LTA$ but is still exponential-like \cite{spectrum}. 
	
	Decreasing monomial codes are a class of codes with algebraic structure, including polar codes and Reed-Muller (RM) codes. Although it is difficult to obtain the complete code spectrum of decreasing monomial codes, in reality, the code spectrum of the low weight codewords is sufficient to reflect the ML performance of the codes. For the research of RM codes, in \cite{RM1}, codewords within $2$ times the minimum code weight were well characterized and represented. Subsequently, \cite{RM2} extended the results to codewords with $2.5$ times the minimum code weight. The code spectrum of the low weight codewords in polar codes have also been widely studied recently. In \cite{algebra}, the theoretical characterization of the minimum weight code word count is given based on the algebraic properties of polar codes. Subsequently, \cite{2d1} \cite{2d2} can obtain some low weight codeword weight distributions in polynomial time. Therefore, the study of partial weight code spectrum has been well promoted, but theoretically, there is still no further progress in reduce the complexity of complete code spectrum for polar codes.
	
	In this paper, we further improve the method for calculating the complete code spectrum in \cite{spectrum}. We construct a class of new subgroups of LTA by the structure of LTA under the two monomials, and use these subgroups to divide the cosets with the same code distribution to reduce the computational complexity of the complete code spectrum at any code length and rate. Experimental results show that most of our results are several times less complex than the latest results in \cite{spectrum}.
		
	The rest of this paper is organized as follows. In section \ref{section3}, we review polar codes and decreasing monomial codes, as well as the method of dividing the coset by LTA in \cite{spectrum}. In section \ref{section4}, we give the definition of the new subgroups and our main theorem which describes how to use the subgroups to find the cosets with the same weight distribution and reduce complexity. In section \ref{section5}, we compare the computational complexity of our method and the method in \cite{spectrum}. In section \ref{section6}, we draw some conclusions. Finally, there are several lemmas which are proofed and the proof of our main theorem in the appendixes.

	\section{PRELIMINARIES}\label{section3}

	\subsection{Polar codes and polar cosets}
	Let $N=2^n$,$F_N=B_N F_2^{\otimes n}$, where $F_2=\left(\begin{aligned}
& 1& 0\\
& 1& 1\\
\end{aligned}\right)$ and $B_N$ is the bit-reversal permutation matrix. Then a polar code denoted by $\mathcal{P}(N,K)$ can be generated by choosing $K$ rows in $F_N$ as the information bit set $\mathcal{I}$, and other rows are denoted by $\mathcal{F}$. The transmitted code $\boldsymbol{x}$ can be encoded by $\boldsymbol{x}=\boldsymbol{u} F_N$, where only $\boldsymbol{u}(\mathcal{I})$ can carry information while $\boldsymbol{u}(\mathcal{F})=0$.

The code spectrum of a polar code consists of many polar cosets. Therefore, we give the definition of polar cosets above:

\begin{definition}
Given a vector $\boldsymbol{u}_i \in\{0,1\}^{i+1},0\le i\le N-1$, the polar coset of $\boldsymbol{u}_i$ is defined as

\begin{equation}
C_N(\boldsymbol{u}_i)=\{\boldsymbol{u} F_N \mid \boldsymbol{u}=(\boldsymbol{u}_i,\boldsymbol{v}),\boldsymbol{v} \in \{0,1\}^{N-i-1}\}.
\end{equation}
\end{definition}

To facilitate our discussion of polar code spectrum, we need to introduce the definition of weight enumeration function.

\begin{definition}
Given $\boldsymbol{u}_i \in\{0,1\}^{i+1},0\le i\le N-1$, we define the weight enumerating polynomial for polar coset $C_N(\boldsymbol{u}_i)$:
\begin{equation}
A_N(\boldsymbol{u}_i)(X)=\mathop{\sum}_{d=0}^N A_d X^d
\end{equation}

where $A_d$ is the number of the codeword in $C_N(\boldsymbol{u}_i)$ with Hamming weight $d$.
\end{definition}

In \cite{spectrum}, An algorithm for computing a polar coset is proposed with complexity $O(N^2)$. For the convenience of discussion, we review some definitions of parameters in polar codes.

\begin{definition}
\cite{spectrum} Given a polar code $\mathcal{P}(N,K)$ with information  index set $\mathcal{I}$ and frozen index set $\mathcal{F}$. Then we define the last frozen bit index of polar code $\mathcal{P}(N,K)$ as 

\begin{equation}
\tau(\mathcal{P}(N,K))=max\{\mathcal{F}\}
\end{equation}

and the mixing factor of $\mathcal{P}(N,K)$ as 
\begin{equation}
MF(\mathcal{P}(N,K))=\mid\{i\in \mathcal{I}|i<\tau(\mathcal{P}(N,K))\}\mid
\end{equation}
\end{definition}

In order to analysis the code spectrum of polar codes, we firstly need to represent a polar code as the union of some polar cosets.

\begin{proposition}
$\mathcal{P}(N,K)$ can be denoted by
\begin{equation}
\mathcal{P}(N,K)=\mathop{\bigcup}_{\boldsymbol{u}_{\tau}\in \{0,1\}^{\tau+1},\boldsymbol{u}_{\tau}(\mathcal{F})=0}C_N(\boldsymbol{u}_{\tau})
\end{equation}
\end{proposition}
	
	\subsection{Decreasing monomial codes}

Polar codes are a class of decreasing monomial codes [2], so we give some definitions of decreasing monomial codes for later use.

Let ring $R_n = \mathbb{F}_2[x_0,\cdots,x_{n-1}]/(x_0^2-x_0,\cdots,x_{n-1}^2-x_{n-1})$. Given $g\in R_n$, the homomorphism of algebra $ev:R_n\rightarrow \{0,1\}^N$ is defined as

\begin{equation}
ev(g)=\left(g(\boldsymbol{u})\right)_{\boldsymbol{u}\in \mathbb{F}_2^n}
\end{equation}
	
	The set of monomial is defined as

\begin{equation}
\mathcal{M}_n=\{x_0^{a_0} \cdots  x_{n-1}^{a_{n-1}}\mid (a_0,\cdots,a_{n-1})\in \mathbb{F}_2^n\}
\end{equation}

\begin{definition}
Let $N=2^n$ and $\mathcal{I}$ is a subset of monomial. Then the monomial code $C(I)$ generated by $\mathcal{I}$ is defined as
\begin{equation}
C(I)=span\{ev(g)\mid g\in \mathcal{I}\}
\end{equation}
\end{definition}
	
Given monomial $g\in \mathcal{M}_n$, let $[g]$ be the row index of $g$ in $F_N$ which will be used frequently.

Before we give the definition of decreasing monomial codes, the partial order of two monomial \cite{alge} need to be introduced.	
	
\begin{definition}
Given two monomials with the same degree, $g=x_{i_1} \cdots x_{i_s},h=x_{j_1} \cdots x_{j_t},i_1<\cdots<i_s,j_1<\cdots<j_s$, then $g\preceq h \Leftrightarrow \forall q\in\{1,\cdots,s\},i_q\le j_q$.

If $g,h$ have different degrees, then we say $g\preceq h \Leftrightarrow \exists $ a divisor $h^\prime$ with of $h$ the same degree as $g$ s.t. $g\preceq h^\prime$.
\end{definition}	
	
	\begin{definition}
	(Decreasing monomial codes \cite{alge}) A code $C(\mathcal{I})$ is a decreasing monomial code, if $\forall h\in \mathcal{I},g\preceq h \Rightarrow g\in \mathcal{I}$.
	\end{definition}
	
    \subsection{LTA and its subgroup}
    Let the affine transformation $(\boldsymbol{A},\boldsymbol{b}):(\boldsymbol{A},\boldsymbol{b})\cdot x=\boldsymbol{A}x+\boldsymbol{b}$, where matrix $\boldsymbol{A}\in \mathbb{F}_2^{n\times n}$, vector $\boldsymbol{b}\in \mathbb{F}_2^{n}$.
    
    \begin{definition}
    All the lower triangular affine transformation $(\boldsymbol{A},\boldsymbol{b})$ form the lower triangular affine group,denoted by $LTA(n,2)$, where the lower triangular affine transformation means $\boldsymbol{A}$ is a lower triangular with $a_{i,i}=1$.
    \end{definition}
    
    Now we review the subgroup of LTA related to a monomial \cite{alge}:
    
    \begin{definition}
    Given $h\in \mathcal{M}_n$,the the subgroup $LTA(n,2)_h$ of group $LTA(n,2)$ is defined as 
    
    \begin{equation}
    (\boldsymbol{A},\boldsymbol{b})\ with\ \left\{    
    \begin{aligned}
    & a_{i,j}=0,if\ i \notin ind(h)\ or\ j\in ind(h)\\
    & b_i=0,if\ i\notin ind(h)\\
    \end{aligned}\right.
    \end{equation}
    
    where $ind(h)$ denote the index which appears in $h$.
    \end{definition}
   
\subsection{Reduce complexity using the group action}   
   In \cite{spectrum}, the authors propose a significant method to reduce the complexity of computing the complete code spectrum for polar codes. And this method inspires our work. Before we introduce the main conclusion in their paper, we review a ovd relation between two monomials defined in \cite{spectrum}:
   
   \begin{definition}
   For two monomials $g,h$, we say $h$ is a one-variable descendant (ovd) of $g$, denoted by $h \prec_o g$ if $g=hx_i$ or $g=qx_i,h=qx_j$ for some monomial $q$ with $i>j$.
   \end{definition}
   
   \begin{example}
   When $n=3$, $x_0x_1 \prec_o x_0 x_2,x_0 x_2 \prec_o x_0 x_1 x_2$. But $x_0$ and $x_0 x_1 x_2$ do not have this ovd relation, because there are two different variables. 
   \end{example}
   
   Use ovd relation, computing the weight distribution of polar code can be simplified by the theorem above \cite{spectrum}.
   
   \begin{theorem}\label{ltacom}
   Given a decreasing monomial code $C(\mathcal{I})$, and let $g$ be the first information bit of $\mathcal{I}$. Then $\mathcal{I}$ can be divided into three parts:
   
   \begin{equation}
   \mathcal{I}=\{g\} \bigcup \mathcal{A} \bigcup \mathcal{B} 
   \end{equation}
   
   where 
   \begin{equation}
   \mathcal{A}=\{f\in \mathcal{I}\mid f\prec_o g, [f]<\tau(C(\mathcal{I}))\}
   \end{equation}
   and
   \begin{equation}
   \mathcal{B}=\mathcal{I}\backslash \left(\{g\} \bigcup \mathcal{A}\right)
   \end{equation}
   
   Then the group action of $LTA(n,2)_g$ on the set $\mathcal{S}$
is transitive,where
\begin{equation}
\mathcal{S}=\{ev(g)+\mathop{\sum}_{h\in \mathcal{B}} a_h\cdot ev(h)+\mathcal{C}(\mathcal{B})|h\in \mathcal{A},a_h\in \{0,1\}\}
\end{equation}
   \end{theorem}
   
   Theorem \ref{ltacom} divides all the $2^{MF(C(\mathcal{I}))}$ cosets into $MF(C(\mathcal{I}))$ parts. And in each part, using the ovd relation $\prec_o$ can find many cosets with the same weight distribution.
   
    \section{An enhanced version in reducing complexity by a new subgroup of LTA }\label{section4} 
     In this chapter, we first provide a definition of a class of subgroups of LTA based on two monomials. Then, based on our proposed subgroups, we introduce our main theorems, which states that our subgroups can be further utilized to find more equivalent coset. Finally, we provide an example to explain the application and benefits of our proposed new subgroups of LTA.
     
   We firstly give the definition of the subgroups for two monomials:  
   
   \subsection{A new subgroup of LTA}
   
Firstly, give two monomials $h_1$, $h_2$ satisfying $[h_1]<[h_2]$ and they have not ovd relation, we define some symbol markings:

We only consider the information bits in the front of the last frozen bits, because only these positions affect the complexity.

$\mathcal{O}_{h_1}$ consists of the ovd of $h_1$ after $h_2$:

\begin{equation}
\mathcal{O}_{h_1}=\{f\in \mathcal{I}\mid h_1\ s.t. f\prec_o h_1,[h_2]<[f]<\tau(C(\mathcal{I}))\}
\end{equation}

$\mathcal{O}_{h_2}$ consists of the ovd of $h_2$:

\begin{equation}
\mathcal{O}_{h_2}=\{f\in \mathcal{I}\mid h_2\ s.t. f\prec_o h_2,[f]<\tau(C(\mathcal{I}))\}
\end{equation}

$\mathcal{O}^c_{h_1}$ consists of monomials after $h_2$ but not the ovd of $h_1$:

\begin{equation}
\mathcal{O}^c_{h_1}=\{f\notin \mathcal{O}_{h_1}\mid [h_2]<[f]<\tau(C(\mathcal{I}))\}
\end{equation}

$\mathcal{O}^c_{h_2}$ consists of monomials after $h_2$ but not the ovd of $h_2$:

\begin{equation}
\mathcal{O}^c_{h_2}=\{f\notin \mathcal{O}_{h_2}\mid [h_2]<[f]<\tau(C(\mathcal{I}))\}
\end{equation}

With the symbol markings, we give the definition of the subgroups of LTA for two monomials:

\begin{definition}
For monomials $h_1$ and $h_2$, we define $LTA(n,2)_{h_2}^{h_1}$ s.t. the elements in matrix $\boldsymbol{A}$ and vector $\boldsymbol{b}$ are equal to zero except:

\begin{equation}
\forall f\in \mathcal{O}_{h_1}\backslash \mathcal{O}_{h_2}\backslash \mathcal{D},\left\{    
    \begin{aligned}
    & a_{s,t}\in \{0,1\},if\ h_1=qx_s,f=qx_t,t<s\\
    & b_r\in \{0,1\},if\ h_1=fx_r\\
    \end{aligned}\right.
\end{equation}

\begin{equation}
\forall f\in \mathcal{O}_{h_2}\backslash \mathcal{O}_{h_1}\backslash \mathcal{D},\left\{    
    \begin{aligned}
    & a_{s,t}\in \{0,1\},if\ h_2=qx_s,f=qx_t,t<s\\
    & b_r\in \{0,1\},if\ h_2=fx_r\\
    \end{aligned}\right.
\end{equation}

where 
   
   \begin{equation}
   \mathcal{D}=\{f\in \mathcal{O}_{h_1}\bigcup \mathcal{O}_{h_2}\mid \left(\mathop{\prod}_{j\in ind(h_1) \cap ind(h_2)}x_j \right) \nmid f\}
   \end{equation}
\end{definition}

\begin{remark}
Intuitively, the group we define is a subgroup of LTA, whose matrix $A$ and vector $b$ contains non-zero positions in the coefficient terms to the ovd of two monomials $h_1$ and $h_2$, but can not contain the coefficient terms to $\mathcal{O}_{h_1} \bigcap \mathcal{O}_{h_2}$ and $\mathcal{D}$. The main reason is that the ovd of $h_1$ has already been utilized in \cite{spectrum}, while our main contribution lies in utilizing the ovd of $h_2$, which involves the connection between the ovd of two monomial. And the role of $\mathcal{D}$ is to ensure that the two mapping processes of $\mathcal{O}_{h_1}\backslash \mathcal{O}_{h_2}\backslash \mathcal{D}$ and $\mathcal{O}_{h_2}\backslash \mathcal{O}_{h_1}\backslash \mathcal{D}$ are independent, so that the initial coset can fill the entire space later.
\end{remark}

We find that our conclusion does not apply to all two monomials, so we limited the scope of the two monomials. On the one hand, this makes our conclusion provable, and on the other hand, in the process of simplifying the code spectrum, our simulation calculations show that the condition of this feasible region is relatively broad, because most of the two monomials that can be further reduced fall within our feasible region.

\begin{definition}
We define the feasible region for $h_1$ and $h_2$:
\begin{equation}
\Omega_{h_1,h_2}=\{(h_1,h_2)\mid \mid h_1\backslash h_2 \mid\le 2\ and\ \mid h_2\backslash h_1 \mid\le 2\}
\end{equation}
\end{definition}

\subsection{The main theorem}

We first need to obtain the following transitional theorem before we can use it to derive the final main theorem.

\begin{theorem}\label{ltalemma}
   Given a decreasing monomial code $C(\mathcal{I})$. Let $h_1$ is the first information set of $\mathcal{I}$. And $h_2$ be the first information set of $\mathcal{O}_{h_1}^c$ with $[h_1]<[h_2],(h_1,h_2)\in \Omega_{h_1,h_2}$. With the same conditions of Theorem \ref{ltaour}, let $\mathcal{W}^{\prime}\subset \mathcal{W}$ be a arbitrary subset, where $\mathcal{W}$ is defined in Appendix \ref{findW} with $\mid \mathcal{W}\mid\le 1$:
   
\begin{equation}
\mathcal{W}=\{g\in \mathcal{O}_{h_1}^c \cap \mathcal{O}_{h_2}^c\mid \exists\ p\in \mathcal{O}_{h_1}\backslash \mathcal{O}_{h_2}\backslash \mathcal{D}\ s.t.\ <(\boldsymbol{A},\boldsymbol{b}) \cdot g>_{p}\ne 0\}
\end{equation}  
   
Then for set $\mathcal{S}_1(\mathcal{W}^{\prime})$ defined by
   
\begin{equation}
\begin{aligned}
\mathcal{S}_1(\mathcal{W}^{\prime})=& \{ev(h_1)+ev(h_2)+\mathop{\sum}_{f\in \mathcal{W}^\prime}ev(f)\\
& +\mathop{\sum}_{h\in \left\{\mathcal{O}_{h_2} \backslash \mathcal{O}_{h_1} \backslash \mathcal{D}\right\} \bigcup \{\mathcal{O}_{h_1}\backslash \mathcal{O}_{h_2}\backslash \mathcal{D}\}} a_h\cdot ev(h)\\
& +\mathcal{C}(\mathcal{B}\backslash \mathcal{W})|\forall h\in \mathcal{A} \backslash \mathcal{D},a_h\in \{0,1\}\}\\
\end{aligned}
\end{equation}

The group action of subgroup $LTA(n,2)_{h_2}^{h_1}$ on the set $\mathcal{S}_1(\mathcal{W}^{\prime})$ is transitive. Therefore,all the cosets in $\mathcal{S}_1(\mathcal{W}^{\prime})$ have the same weight
distribution.
\end{theorem}

Our main theorem is described as follows, which can be used to further reduce the complexity of computing code spectrum.

\begin{theorem}\label{ltaour}
   Given a decreasing monomial code $C(\mathcal{I})$. Let $h_1$ is the first information set of $\mathcal{I}$. And $h_2$ be the first information bit that satisfies the following conditions: $[h_1]<[h_2],h_2\notin \mathcal{O}_{h_1}$,and $(h_1,h_2)\in \Omega_{h_1,h_2}$. Let $\mathcal{I}_1=\mathcal{I}\backslash \{i\in \mathcal{I} \mid i<[h_2],i\ne [h_1]\}$, then $\mathcal{I}_1$ denotes the information sets which only contain $h_1$ and the positions behind $[h_2]$. It can be divided into several parts:
   
   \begin{equation}
    \begin{aligned}
    \mathcal{I}_1= & \{h_1\} \bigcup \{h_2\} \bigcup (\mathcal{O}_{h_1}\backslash \mathcal{O}_{h_2}\backslash \mathcal{D}) \bigcup (\mathcal{O}_{h_2}\backslash \mathcal{O}_{h_1}\backslash \mathcal{D})\\
    & \bigcup (\mathcal{O}_{h_1}\bigcap \mathcal{O}_{h_2})\bigcup (\mathcal{O}_{h_1}\backslash \mathcal{O}_{h_2}\bigcap \mathcal{D}) \bigcup \left\{\mathcal{B}\right\}\\
    \end{aligned}    
   \end{equation}
   
   where
   \begin{equation}
   \mathcal{B}= (\mathcal{O}_{h_2} \backslash \mathcal{O}_{h_1}\bigcap \mathcal{D}) \bigcup \left(\mathcal{O}_{h_1}^c\bigcap \mathcal{O}_{h_2}^c \right)\bigcup\mathcal{T}
   \end{equation}
   
   where $\mathcal{T}$ is the information positions behind $\tau(C(\mathcal{I}))$.
   
   Then for set $\mathcal{S}$ defined by
\begin{equation}
\begin{aligned}
\mathcal{S}= \{ev(h_1)+ev(h_2)+& \mathop{\sum}_{h\in \left\{\mathcal{O}_{h_2}\backslash \mathcal{O}_{h_1} \backslash \mathcal{D}\right\} \bigcup \{\mathcal{O}_{h_1}\backslash \mathcal{O}_{h_2}\backslash \mathcal{D}\}} a_h\cdot ev(h)\\
&  +\mathcal{C}(\mathcal{B})|\forall h\in \mathcal{A} \backslash \mathcal{D},a_h\in \{0,1\}\}\\
\end{aligned}
\end{equation}
\end{theorem}

all the cosets of $\mathcal{C}(\mathcal{B})$ in $\mathcal{S}$ have the same weight
distribution.

\subsection{An example}

We review the example in \cite{spectrum} and show the influence of our method on the example. For a polar code $\mathcal{P}(32,24)$ and we choose the information bit set by the method in 5G \cite{5G}.

In the first monomial $x_0 x_3 x_4$, their method compute $16$ cosets.

If we compute twice again, then we only need to compute $6$ cosets. In detail,

After we do the first time like \cite{spectrum}, we get four monomials $x_1x_2x_4,x_2x_4,x_1x_4,x_4 \in \mathcal{O}^c_{x_0 x_3 x_4}$.

We do the second time for the 16 cosets:

For $h_1=x_0x_3x_4$, 

\begin{equation}
\begin{aligned}
\mathbb{C}(1a_{x_1x_2x_4}a_{x_2x_4}& a_{x_1x_4}a_{x_4})=\mathbb{C}(10000)+\mathbb{C}(11a_{x_2x_4}a_{x_1x_4}a_{x_4})\\
& +\mathbb{C}(101a_{x_1x_4}a_{x_4})+\mathbb{C}(1001a_{x_4})+\mathbb{C}(10001)\\
\end{aligned}
\end{equation}

$\bullet$ We firstly see that how the new subgroup effects on the second monomial $h_2=x_1x_2 x_4$, we consider $\mathbb{C}(11a_{x_2x_4}a_{x_1x_4}a_{x_4})$ where $x_2x_4$, $x_1x_4$ both are the ovd of $x_1x_2 x_4$, and they both contain $\mathop{\prod}_{i\in ind(h_1)\cap ind(h_2)} x_i=x_4$, so their weight distribution are the same by Theorem \ref{ltalemma}. Therefore, $\mathbb{C}(11a_{x_2x_4}a_{x_1x_4}a_{x_4})=2^2 \cdot \mathbb{C}(1100a_{x_4})$

Other cases are similar, so we can simplify the coset by 

\begin{equation}
\begin{aligned}
\mathbb{C}(1a_{x_1x_2x_4}a_{x_2x_4}& a_{x_1x_4}a_{x_4})=\mathbb{C}(10000)+2^2 \cdot \mathbb{C}(1100a_{x_4})\\
& +2^2 \cdot \mathbb{C}(10100)+2 \mathbb{C}(10010)+\mathbb{C}(10001)\\
\end{aligned}
\end{equation}

Finally, we reduce the complexity of the first monomial from $16$ to $6$. And for the whole code, we can reduce the complexity from $40$ to $24$.

$\bullet$ Then in detail, we see the algebraic structure of the new subgroup for $h_2=x_1x_2 x_4$. We can see the subgroup is $LTA(n,2)_{x_1x_2x_4}^{x_0x_3x_4}$. Because $x_1x_4,x_2x_4\in \mathcal{O}_{h_2}\backslash \mathcal{O}_{h_1}\backslash \mathcal{D}$ and $x_0x_4\in \mathcal{O}_{h_1}\backslash \mathcal{O}_{h_2}\backslash \mathcal{D}$, so according to our definition, $\forall (\boldsymbol{A},\boldsymbol{b}) \in LTA(n,2)_{x_1x_2x_4}^{x_0x_3x_4}$, the subgroup of LTA has the form above:

$$
A=\left(
\begin{array}{lllll}
1& 0& 0& 0& 0\\
0& 1& 0& 0& 0\\
0& 0& 1& 0& 0\\
0& 0& 0& 1& 0\\
0& 0& 0& 0& 1\\
\end{array}
 \right),b=\left(
\begin{array}{l}
0\\
b_1\\
b_2\\
b_3\\
0\\
\end{array}
 \right)
$$

We can see $\mathcal{W}=\emptyset$ and 

\begin{equation}
\begin{aligned}
& (\boldsymbol{A},\boldsymbol{b}) \cdot (ev(h_1)+ev(h_2)+C(\mathcal{B}\backslash \mathcal{W}))\\
=& ev(x_0x_3x_4)+ev(x_1x_2x_4)+b_1\cdot ev(x_2x_4)\\
& +b_2\cdot ev(x_1x_4)+b_3\cdot ev(x_0x_4)+C(\mathcal{B}\backslash \mathcal{W})\\
\end{aligned}
\end{equation}

And we only need $b_3=0$ for the twice step. This is because $x_0x_4 \in \mathcal{O}_{h_1}$ have been used for the first monomial $h_1$.

	\section{Applications of our enhanced method}\label{section5}
	In this section, we give some examples for applications and comparisons. Table \ref{table 1} and Table \ref{table 2} give several cases for $N=128$ and $N=256$ relatively with reliability construction in 5G \cite{5G}. The third column is the complexity of the method in \cite{spectrum}. The fourth column is the complexity of our method. And the last column is the comparison of complexity between two methods: $\frac{N_{ovd}}{N_{enhanced}}$, where $N_{ovd}$ and $N_{enhanced}$ denote the number of cosets which need to be computed of two methods. We can see that the degree of complexity reduction varies in different cases, with the highest reduction of about $44$ times among the given cases. Table \ref{table 3} gives the whole weight distribution for $\mathcal{P}(128,32)$ from our algorithm. In this case, our method has reduced the complexity by about $11$ times compared to the latest result.

	\begin{table}[htbp]
		\caption{The complexity of weight distribution for $N=128$}
		\begin{center}
\begin{tabular}{|r|l|r|r|r|}
\hline $\mathcal{P}(N,K)$ & $2^{|MF(\mathbb{C})|}$ & $N_{ovd} $ \cite{spectrum} & $N_{enhanced}$& $\frac{N_{ovd}}{N_{enhanced}}$\\
\hline $\mathcal{P}(128,21)$ & $2^{14}$ &$1446$& $235$ &  $6.15$\\
\hline $\mathcal{P}(128,32)$ & $2^{17}$ &$5292$& $464$ &  $11.41$\\
\hline $\mathcal{P}(128,35)$ & $2^{20}$ &$21336$& $3118$ &  $6.84$\\
\hline $\mathcal{P}(128,41)$ & $2^{26}$ &$2^{19.38}$& $2^{17.26}$ &  $4.37$\\
\hline $\mathcal{P}(128,50)$ & $2^{27}$ &$2^{20.54}$& $2^{19.78}$ &  $1.70$\\
\hline $\mathcal{P}(128,57)$ & $2^{30}$ &$2^{23.02}$& $2^{20.75}$ &  $4.81$\\
\hline $\mathcal{P}(128,64)$ & $2^{34}$ &$2^{25.23}$& $2^{22.51}$ & $6.59$\\
\hline $\mathcal{P}(128,69)$ & $2^{38}$ &$2^{27.05}$& $2^{23.86}$ &  $9.12$\\
\hline $\mathcal{P}(128,73)$ & $2^{42}$ &$2^{30.22}$& $2^{27.79}$ & $5.39$\\
\hline $\mathcal{P}(128,82)$ & $2^{51}$ &$2^{38.42}$& $2^{37.41}$ &  $2.02$\\
\hline $\mathcal{P}(128,90)$ & $2^{43}$ &$2^{33.76}$& $2^{32.81}$ &  $1.93$\\
\hline $\mathcal{P}(128,98)$ & $2^{39}$ &$2^{30.00}$& $2^{28.11}$ &  $3.72$\\
\hline $\mathcal{P}(128,107)$ & $2^{45}$ &$2^{34.25}$& $2^{31.53}$ &  $6.59$\\
\hline
\end{tabular}
\label{table 1}
		\end{center}
	\end{table}

	\begin{table}[htbp]
		\caption{The complexity of weight distribution for $N=256$}
		\begin{center}
\begin{tabular}{|r|l|r|r|r|}
\hline $\mathcal{P}(N,K)$ & $2^{|MF(\mathbb{C})|}$ & $N_{ovd} $ \cite{spectrum} & $N_{enhanced}$ & $\frac{N_{ovd}}{N_{enhanced}}$\\
\hline $\mathcal{P}(256,44)$ & $2^{29}$ &$2^{23.35}$& $2^{17.88}$ &  $44.07$\\
\hline $\mathcal{P}(256,64)$ & $2^{37}$ &$2^{29.48}$& $2^{28.62}$ &  $1.82$\\
\hline $\mathcal{P}(256,85)$ & $2^{54}$ &$2^{44.67}$& $2^{42.19}$ &  $5.58$\\
\hline $\mathcal{P}(256,107)$ & $2^{76}$ &$2^{62.72}$& $2^{59.08}$ &  $12.47$\\
\hline $\mathcal{P}(256,126)$ & $2^{79}$ &$2^{64.74}$& $2^{61.83}$  & $7.51$\\
\hline $\mathcal{P}(256,157)$ & $2^{95}$ &$2^{79.25}$& $2^{78.02}$  & $2.34$\\
\hline $\mathcal{P}(256,172)$ & $2^{109}$ &$2^{92.63}$& $2^{89.99}$ &  $6.25$\\
\hline $\mathcal{P}(256,179)$ & $2^{116}$ &$2^{99.36}$& $2^{95.58}$ &  $13.82$\\
\hline $\mathcal{P}(256,185)$ & $2^{122}$ &$2^{104.77}$ &$2^{100.28}$  &  $22.47$\\
\hline $\mathcal{P}(256,192)$ & $2^{129}$ &$2^{111.16}$ &$2^{107.73}$  &  $10.78$\\
\hline
\end{tabular}
\label{table 2}
		\end{center}
	\end{table}
	
		\begin{table}[htbp]
		\caption{The whole weight distribution for $N=128,K=32$ from our algorithm}
		\begin{center}
\begin{tabular}{|r|r|}
\hline $d$ & $A_d$\\
\hline 0 &1\\
\hline 16 &152\\
\hline 24 &384\\
\hline 32 &40540\\
\hline 40 &824192\\
\hline 48 &33530152\\
\hline 52 &88080384\\
\hline 56 &579218176\\
\hline 60 &616562688\\
\hline 64 &1658453958\\
\hline 68 &616562688\\
\hline 72 &579218176\\
\hline 76 &88080384\\
\hline 80 &33530152\\
\hline 88 &824192\\
\hline 96 &40540\\
\hline 104 &384\\
\hline 112 &152\\
\hline 128 &1\\
\hline
\end{tabular}
\label{table 3}
		\end{center}
	\end{table}

	\section{Conclusion}\label{section6}
	
	In this paper, we enhance the deterministic algorithm \cite{spectrum} by defining and utilizing a new subgroup of LTA. We find more cosets with the same weight distribution and even some cosets are permuted by two different permutations. Our enhanced algorithm can compute the spectrum code length $128$ with a faster speed. Actually we observed the real number of different cosets are few, so there may exist a larger group to do this. But the key challenges is how to find the algebraic structures related to codes like LTA.

\begin{appendices}
\section{the group $LTA(n,2)_{h_2}^{h_1}$ actions on $h_1$}

Due to the fact that any term $x_i$ in a monomial maps to another monomial, either through coefficient $a_{ij}$ for some $j$ or coefficient $b_i$, for simplicity, we unify the coefficient of the mapping process as $e_i$. It denotes the mapping process by LTA. We use $e_{\mathcal{O}_{h_i}}$ to represent the mapping process from $h_i$ to $\mathcal{O}_{h_i}$, $i=1,2$. The subgroup we define has only two mapping processes: $e_{\mathcal{O}_{h_1}\backslash \mathcal{O}_{h_2}\backslash \mathcal{D}}$ and $e_{\mathcal{O}_{h_2}\backslash \mathcal{O}_{h_1}\backslash \mathcal{D}}$, where the role of $\mathcal{D}$ is to ensure that the two mapping processes are independent. This ensures that the mononomial $h_2$ can further reduce complexity based on the monomial $h_1$.

\begin{lemma}\label{action on h1}
With the same conditions of Theorem \ref{ltaour},for any $(\boldsymbol{A},\boldsymbol{b})\in LTA(n,2)_{h_2}^{h_1}$,

\begin{equation}
\left\{
\begin{aligned}
&<(\boldsymbol{A},\boldsymbol{b}) \cdot h_1>_{\mathcal{O}_{h_1}\backslash \mathcal{O}_{h_2}\backslash \mathcal{D}} =e_{\mathcal{O}_{h_1}\backslash \mathcal{O}_{h_2}\backslash \mathcal{D}} \\
&<(\boldsymbol{A},\boldsymbol{b}) \cdot h_1>_{\mathcal{O}_{h_1}\backslash \mathcal{O}_{h_2}\bigcap \mathcal{D}} =0 \\
& <(\boldsymbol{A},\boldsymbol{b}) \cdot h_1>_{\mathcal{O}_{h_1}\bigcap \mathcal{O}_{h_2}} =0\\
&<(\boldsymbol{A},\boldsymbol{b}) \cdot h_1>_{h_2} =0 \\
& <(\boldsymbol{A},\boldsymbol{b}) \cdot h_1>_{\mathcal{O}_{h_2}\backslash \mathcal{O}_{h_1}\backslash \mathcal{D}} =0\\
\end{aligned}\right.
\end{equation}

\end{lemma}

\pf 

$\bullet<(\boldsymbol{A},\boldsymbol{b}) \cdot h_1>_{\mathcal{O}_{h_1}\backslash \mathcal{O}_{h_2}\backslash \mathcal{D}} =e_{\mathcal{O}_{h_1}\backslash \mathcal{O}_{h_2}\backslash \mathcal{D}},<(\boldsymbol{A},\boldsymbol{b}) \cdot h_1>_{\mathcal{O}_{h_1}\backslash \mathcal{O}_{h_2}\bigcap \mathcal{D}} =0$ and $<(\boldsymbol{A},\boldsymbol{b}) \cdot h_1>_{\mathcal{O}_{h_1}\bigcap \mathcal{O}_{h_2}} =0$ are obvious in the definition of group.

$\bullet <(\boldsymbol{A},\boldsymbol{b}) \cdot h_1>_{h_2} =0$:

(i) If $h_1\backslash h_2 =\{i\}$, then 

(a) $| h_2\backslash h_1 |\le 1$: $h_1=h_2 x_i$ or $h_1=\frac{h_2 x_i}{x_j}$ for some $x_j$. This is a contradiction because $h_2\notin \mathcal{O}_{h_1}$; 

(b) $| h_2\backslash h_1 |\ge 2$:$deg(h_2)>deg(h_1)$, so $<(\boldsymbol{A},\boldsymbol{b}) \cdot h_1>_{h_2} =0$;

(ii) If $| h_1\backslash h_2 | \ge 2$, then for $i \in ind(h_1)\backslash ind(h_2)$ with $i\ne max(ind(h_1)\backslash ind(h_2))$, we have $e_i=0$ because $[\frac{h_1 x_j}{x_i}]<[\frac{h_1}{x_i}]<[h_2] (\forall j)$. So $<(\boldsymbol{A},\boldsymbol{b}) \cdot h_1>_{h_2} =0$.

$\bullet <(\boldsymbol{A},\boldsymbol{b}) \cdot h_1>_{\mathcal{O}_{h_2}\backslash \mathcal{O}_{h_1}\backslash \mathcal{D}} =0:$

For $f\in \mathcal{O}_{h_2}\backslash \mathcal{O}_{h_1}\backslash \mathcal{D}$,

(i) $h_2=fx_s,x_s\in h_2\backslash h_1$:

If $h_1\backslash h_2 =\{i\}$, then 

(a) $| h_2\backslash h_1 |\le 1$: $h_1=h_2 x_i$ or $h_1=\frac{h_2 x_i}{x_s}$. This is a contradiction because $h_2\notin \mathcal{O}_{h_1}$; 

(b) $| h_2\backslash h_1 |=2$: $f=\frac{h_1 x_j}{x_i}\in \mathcal{O}_{h_1}$ which is a contradiction;

(c) $| h_2\backslash h_1 |\ge 3$:$deg(f)>deg(h_1)$, so $<(\boldsymbol{A},\boldsymbol{b}) \cdot h_1>_{f} =0$;

If $| h_1\backslash h_2 | \ge 2$, samely, for $i\in ind(h_1)\backslash ind(h_2)$ with $i\ne max(ind(h_1)\backslash ind(h_2))$, we have $e_i=0$ because $[\frac{h_1 x_j}{x_i}]<[\frac{h_1}{x_i}]<[h_2] (\forall j)$. So $<(\boldsymbol{A},\boldsymbol{b}) \cdot h_1>_{f} =0$.

(ii) $h_2=qx_{s_1}$, $f=qx_{s_2}$, $x_{s_1}\in h_2\backslash h_1$:

If $h_1\backslash h_2 =\{i\}$, then 

(a) $| h_2\backslash h_1 |\le 1$: $h_1=h_2 x_i$ or $h_1=\frac{h_2 x_i}{x_{s_1}}$. This is a contradiction because $h_2\notin \mathcal{O}_{h_1}$; 

(b) $| h_2\backslash h_1 |\ge 2$: $deg(f)>deg(h_1)$, so $<(\boldsymbol{A},\boldsymbol{b}) \cdot h_1>_{f} =0$;

If $| h_1\backslash h_2|=2$, $x_{s_2}\in h_1\backslash h_2,x_{s_2}\ne max\{inf(h_1\backslash h_2)\}$. $| h_1\backslash f |=1$,

and 
\begin{equation}
\left\{
\begin{aligned}
& if\ | f\backslash h_1 |\le 1,f\in \mathcal{O}_{h_1}\ which\ is\ a\ contradiction;\\
& if\ | f\backslash h_1 |> 1,deg(f)>deg(h_1),so\ <(\boldsymbol{A},\boldsymbol{b}) \cdot h_1>_{f} =0.\\
\end{aligned}\right.
\end{equation}

Else if $| h_1\backslash h_2 | \ge 2$, samely, for $i\in ind(h_1)\backslash ind(h_2)$ with $i\ne max(ind(h_1)\backslash ind(h_2))$, we have $e_i=0$ because $[\frac{h_1 x_j}{x_i}]<[\frac{h_1}{x_i}]<[h_2] (\forall j)$. So $<(\boldsymbol{A},\boldsymbol{b}) \cdot h_1>_{f} =0$.

$\bullet <(\boldsymbol{A},\boldsymbol{b}) \cdot h_1>_{\mathcal{O}_{h_1}\bigcap \mathcal{O}_{h_2}} =0$: By the definition of group.

\e
\section{the group $LTA(n,2)_{h_2}^{h_1}$ actions on $h_2$}
\begin{lemma}\label{action on h2}
With the same conditions of Theorem \ref{ltaour}, for any $(\boldsymbol{A},\boldsymbol{b})\in LTA(n,2)_{h_2}^{h_1}$,

\begin{equation}
\left\{
\begin{aligned}
&<(\boldsymbol{A},\boldsymbol{b}) \cdot h_2>_{\mathcal{O}_{h_2}\backslash \mathcal{O}_{h_1}\backslash \mathcal{D}} =e_{\mathcal{O}_{h_2}\backslash \mathcal{O}_{h_1}\backslash \mathcal{D}} \\
&<(\boldsymbol{A},\boldsymbol{b}) \cdot h_2>_{\mathcal{O}_{h_1}\backslash \mathcal{O}_{h_2}\backslash \mathcal{D}} =0\ or\ a\ function\ of\ e_{\mathcal{O}_{h_2}\backslash \mathcal{O}_{h_1}\backslash \mathcal{D}} \\
&<(\boldsymbol{A},\boldsymbol{b}) \cdot h_2>_{\mathcal{O}_{h_1}\backslash \mathcal{O}_{h_2}\bigcap \mathcal{D}} =0 \\
& <(\boldsymbol{A},\boldsymbol{b}) \cdot h_2>_{\mathcal{O}_{h_2}\backslash \mathcal{O}_{h_1}\bigcap \mathcal{D}} =0\\
& <(\boldsymbol{A},\boldsymbol{b}) \cdot h_2>_{\mathcal{O}_{h_1}\bigcap \mathcal{O}_{h_2}} =0\\
\end{aligned}\right.
\end{equation}
\end{lemma}

\pf

$\bullet <(\boldsymbol{A},\boldsymbol{b}) \cdot h_2>_{\mathcal{O}_{h_1}\backslash \mathcal{O}_{h_2}\bigcap \mathcal{D}} =0:$

(i) Case1:$h_1=fx_s,s\in ind(h_1) \cap ind(h_2)$, then $e_s=0$;

(ii) Case2:$h_1=px_{s_1},f=px_{s_2},s_1 \in ind(h_1) \cap ind(h_2)$, then $e_{s_1}=0$;

$\bullet <(\boldsymbol{A},\boldsymbol{b}) \cdot h_2>_{\mathcal{O}_{h_1}\backslash \mathcal{O}_{h_2}\backslash \mathcal{D}} =0$ or a function of $e_{\mathcal{O}_{h_2}\backslash \mathcal{O}_{h_1}\backslash \mathcal{D}}$:

For $i\in ind(h_2)\backslash ind(h_1)$, we know $\frac{h_2}{x_i},\frac{h_2 x_j}{x_i}\in \mathcal{O}_{h_2}$, so $e_i\ne 0$ only for $e_{\mathcal{O}_{h_2}\backslash \mathcal{O}_{h_1}\backslash \mathcal{D}}$.

Others can be obtained from the definition of group.
\e

\section{the group $LTA(n,2)_{h_2}^{h_1}$ actions on $\mathcal{O}_{h_2}\backslash \mathcal{O}_{h_1}\bigcap \mathcal{D}$}

\begin{lemma}\label{action on Oh2}
With the same conditions of Theorem \ref{ltaour}, for any $(\boldsymbol{A},\boldsymbol{b})\in LTA(n,2)_{h_2}^{h_1}$,

\begin{equation}
\left\{
\begin{aligned}
<(\boldsymbol{A},\boldsymbol{b}) \cdot f>_{\mathcal{O}_{h_2}\backslash \{f\}} =0,& \forall f\in \mathcal{O}_{h_2} \\
<(\boldsymbol{A},\boldsymbol{b}) \cdot f>_{\mathcal{O}_{h_1}\backslash \mathcal{O}_{h_2}} =0,& \forall f\in \mathcal{O}_{h_2}\backslash \mathcal{O}_{h_1}\bigcap \mathcal{D},\\
& if\ (h_1,h_2)\in \Omega_{h_1,h_2}\\
\end{aligned}\right.
\end{equation}
\end{lemma}

\pf

$\bullet <(\boldsymbol{A},\boldsymbol{b}) \cdot f>_{\mathcal{O}_{h_2}\backslash \{f\}} =0,\forall f\in \mathcal{O}_{h_2}$:

For $g\in \mathcal{O}_{h_2}\backslash \{f\}$:

(i) $h_2=fx_s$, $h_2=gx_t$, $s,t\in ind(h_2) \Rightarrow a_{ts}=0$;

(ii) $h_2=fx_s$, $h_2=qx_{t_1}$, $g=qx_{t_2},t_2<t_1,deg(g)>deg(f)$;

(iii) $h_2=px_{s_1}$, $f=px_{s_2}$, $s_2<s_1$, $h_2=gx_t$, $g=\frac{fx_{s_1}}{x_{s_2}x_t}$,
\begin{equation}
<(\boldsymbol{A},\boldsymbol{b}) \cdot f>_{g} =a_{ts_1} b_{s_2}+a_{s_2s_1}b_t
\end{equation}
$s_2<s_1 \Rightarrow a_{s_2s_1}=0$ and $t,s_1\in ind(h_2)\Rightarrow a_{ts_1}=0$.

(iv) $h_2=px_{s_1}$, $f=px_{s_2}$, $s_2<s_1$, $h_2=qx_{t_1}$, $g=qx_{t_2}$, $t_2<t_1$, $g=\frac{fx_{s_1}x_{t_2}}{x_{s_2}x_{t_1}}$,
\begin{equation}
<(\boldsymbol{A},\boldsymbol{b}) \cdot f>_{g} =a_{t_1s_1} a_{s_2t_2}+a_{s_2s_1}a_{t_1t_2}
\end{equation}
$s_2<s_1 \Rightarrow a_{s_2s_1}=0$ and $t_1,s_1\in ind(h_2)\Rightarrow a_{t_1 s_1}=0$.

In whole, $ <(\boldsymbol{A},\boldsymbol{b}) \cdot f>_{\mathcal{O}_{h_2}\backslash \{f\}} =0.$

$\bullet <(\boldsymbol{A},\boldsymbol{b}) \cdot f>_{\mathcal{O}_{h_1}\backslash \mathcal{O}_{h_2}} =0$, $\forall f\in \mathcal{O}_{h_2}\backslash \mathcal{O}_{h_1}\bigcap \mathcal{D}$, if $(h_1,h_2)\in \Omega_{h_1,h_2}$: For $g\in \mathcal{O}_{h_1}\backslash \mathcal{O}_{h_2}$,

(i) $h_2=fx_s$, $s\in ind(h_1)$, $h_1=gx_t$

(a) $s=t$: $[g]<[f]$, $<(\boldsymbol{A},\boldsymbol{b}) \cdot f>_{g} =0$;

(b) $s\in ind(g)$: $\forall i\in ind(f)$, $i,s \in ind(h_2) \Rightarrow a_{is}=0$.

(ii) $h_2=fx_s$, $s\in ind(h_1)$, $h_1=qx_{t_1}$, $g=qx_{t_2}$, $t_2<t_1$

(a) $s=t_1$: $[g]<[q]<[f]$, $<(\boldsymbol{A},\boldsymbol{b}) \cdot f>_{g} =0$;

(b) $s\in ind(q)$: $\forall i\in ind(f)$, $i,s \in ind(h_2) \Rightarrow a_{is}=0$.

(iii) $h_2=qx_{s_1}$, $f=qx_{s_2}$, $s_{2}<s_1$, $s_1\in ind(h_1)$, $h_1=gx_t$

(a) $s_1=t$: because $x_{t}=x_{s_1}\in h_1 \cap h_2$, so $max\{ind(h_1\backslash h_2)\}\in ind(g)$. But for $j\in ind(f)$,

\begin{equation}
\left\{\begin{aligned}
& if\ j\in h_1\cap h_2,e_j=0;\\
& if\ j\in h_2\backslash h_1,j<max\{ind(h_2\backslash h_1)\}<max\{ind(h_1\backslash h_2)\}.\\
\end{aligned}\right.
\end{equation}

So $s_2=max\{ind(h_1\backslash h_2)\}$, and according to $(h_1,h_2)\in \Omega_{h_1,h_2}$, we focus on the path from $i\in q \subset f$ and $i\in h_2\backslash h_1$  to $j\in g$.

$\bullet \mid h_1\backslash h_2\mid =1$, $\mid h_2\backslash h_1\mid =1$: $h_2\in \mathcal{O}_{h_1}$ contradicts;

$\bullet \mid h_1\backslash h_2\mid =1$, $\mid h_2\backslash h_1\mid =2$: this case $i$ only generate $\emptyset$ because $g\backslash x_{s_2}$ have no elements in $h_1\backslash h_2$ .But $\frac{h_2}{x_i}\in \mathcal{O}_{h_1} \bigcap \mathcal{O}_{h_2}$, so $e_i=b_i=0$;

$\bullet \mid h_1\backslash h_2\mid =2$, $\mid h_2\backslash h_1\mid =1$: $\frac{h_2 x_j}{x_i}\in \mathcal{O}_{h_1} \bigcap \mathcal{O}_{h_2}$, so $e_i=0$;

$\bullet \mid h_1\backslash h_2\mid =2$, $\mid h_2\backslash h_1\mid =2$: $\frac{h_2 x_j}{x_i}\in \mathcal{O}_{h_1} \bigcap \mathcal{O}_{h_2}$, so $e_i=0$.

(b) $s_1\in g$: then any path from $i\in ind(f)$ to $x_{s_1}$, $a_{is_1}=0$ because $i \in h_2$ or $i=s_2<s_1$.

(iv) $h_2=qx_{s_1}$, $f=qx_{s_2}$, $s_{2}<s_1$, $s_1\in ind(h_1)$, $h_1=px_{t_1}$, $g=px_{t_2}$, $t_2<t_1$

(a) $s_1=t_1$: because $x_{t_1}=x_{s_1}\in h_1 \cap h_2$, so $max\{ind(h_1\backslash h_2)\}\in ind(p)\subset ind(g)$. But for $j\in ind(f)$,

\begin{equation}
\left\{\begin{aligned}
& if\ j\in h_1\cap h_2,e_j=0;\\
& if\ j\in h_2\backslash h_1,j<max\{ind(h_2\backslash h_1)\}<max\{ind(h_1\backslash h_2)\}.\\
\end{aligned}\right.
\end{equation}

So $s_2=max\{ind(h_1\backslash h_2)\}$, and according to $(h_1,h_2)\in \Omega_{h_1,h_2}$, we focus on the path from $i\in q \subset f$ and $i\in h_2\backslash h_1$  to $j\in g$.

$\bullet \mid h_1\backslash h_2\mid =1$, $\mid h_2\backslash h_1\mid =1$: $h_2\in \mathcal{O}_{h_1}$ contradicts;

$\bullet \mid h_1\backslash h_2\mid =1$, $\mid h_2\backslash h_1\mid =2$: this case $i$ generate $\emptyset$ or $x_{s_2}$. We only focus on $i$ which generates $\emptyset$. Then $\frac{h_2}{x_i}\in \mathcal{O}_{h_1} \bigcap \mathcal{O}_{h_2}$, so $e_i=b_i=0$;

$\bullet \mid h_1\backslash h_2\mid =2$, $\mid h_2\backslash h_1\mid =1$: $deg(g)>deg(f)$;

$\bullet \mid h_1\backslash h_2\mid =2$, $\mid h_2\backslash h_1\mid =2$: $\frac{h_2 x_j}{x_i}\in \mathcal{O}_{h_1} \bigcap \mathcal{O}_{h_2}$, so $e_i=0$.

(b) $s_1\in p$: then any path from $i\in ind(f)$ to $x_{s_1}$, $a_{is_1}=0$ because $i\in h_2$ or $i=s_{2}<s_1$, $s_1 \in h_2$.
\e

\section{the group $LTA(n,2)_{h_2}^{h_1}$ actions on $\mathcal{O}^c_{h_1}\bigcap \mathcal{O}^c_{h_2}$}\label{findW}

\begin{lemma}\label{action on Ohc1 Ohc2}
With the same conditions of Theorem \ref{ltaour}, for any $(\boldsymbol{A},\boldsymbol{b})\in LTA(n,2)_{h_2}^{h_1}$, $\forall f\in \mathcal{O}^c_{h_1}\bigcap \mathcal{O}^c_{h_2}$

\begin{equation}
\left\{
\begin{aligned}
&<(\boldsymbol{A},\boldsymbol{b}) \cdot f>_{\mathcal{O}_{h_1}\bigcap \mathcal{O}_{h_2}} =0 \\
&<(\boldsymbol{A},\boldsymbol{b}) \cdot f>_{\mathcal{O}_{h_2}\backslash \mathcal{O}_{h_1}\backslash \mathcal{D}} =0 \\
&<(\boldsymbol{A},\boldsymbol{b}) \cdot f>_{\mathcal{O}_{h_1}\backslash \mathcal{O}_{h_2} \bigcap \mathcal{D}} =0 \\
&<(\boldsymbol{A},\boldsymbol{b}) \cdot f>_{\mathcal{O}_{h_1}\backslash \mathcal{O}_{h_2}\backslash \mathcal{D}} =0\ or\ a\ function\ of\ e_{\mathcal{O}_{h_2}\backslash \mathcal{O}_{h_1}\backslash \mathcal{D}} \\
\end{aligned}\right.
\end{equation}
\end{lemma}

\pf

$\bullet <(\boldsymbol{A},\boldsymbol{b}) \cdot f>_{\mathcal{O}_{h_1}\bigcap \mathcal{O}_{h_2}} =0$:

(i) Case1: $h_1=gx_s$, $h_2=gx_t$:

 then $f\in\{gx_s x_t,gx_s,gx_t,g\}$, this is a contradiction;

(ii) Case2: $h_1=gx_s$, $h_2=qx_{t_1}$, $g=qx_{t_2}$, $t_2<t_1$: then $h_1=qx_{t_2}x_s$.$h_1\cap h_2=q$ (if $h_1\cap h_2=h_2$, $h_2\in \mathcal{O}_{h_1}$ is a contradiction),$h_1\backslash h_2=\{x_{t_2},x_s\}$, $h_2\backslash h_1=x_{t_1}$.

If $x_{t_2}\in f$, then $f\in\{gx_s x_{t_1}$, $gx_s,gx_{t_1},g\}$, this is a contradiction.

If $x_{t_2}\notin f$, then only one path to $t_2$ is from $x_{t_1}$. But $qx_{t_1}=h_2$ which is a contradiction.

(iii) Case3: $h_1=px_{s_1}$, $g=px_{s_2}$, $s_2<s_1$, $h_2=gx_t$: then $h_2=px_{s_2}x_t$. $h_1\cap h_2=p$ (if $h_1\cap h_2=h_1$, $[h_2]<[h_1]$ is a contradiction), $h_1\backslash h_2=x_{s_1}$, $h_2\backslash h_1=\{x_{s_2},x_t\}$.

If $x_{s_2}\in f$, then $f\in\{gx_t x_{s_1},gx_t,gx_{s_1},g\}$, this is a contradiction.

If $x_{s_2}\notin f$, then only one path to $s_2$ is from $x_{s_1}$. But $px_{s_1}=h_1$ which is a contradiction.

(iv) Case4: $h_1=px_{s_1}$, $g=px_{s_2}$, $s_2<s_1$, $h_2=qx_{t_1}$, $g=qx_{t_2}$, $t_2<t_1$: then $h_1\cap h_2=\frac{g}{x_{s_2}x_{t_2}}$$, $$h_1\backslash h_2=\{x_{t_2},x_{s_1}\}$, $h_2\backslash h_1=\{x_{s_2},x_{t_1}\}$.

If $x_{s_2},x_{t_2}\in f$, then $f\in\{gx_{t_1} x_{s_1},gx_{t_1},gx_{s_1},g\}$, this is a contradiction.

If $x_{s_2}\notin f,x_{t_2}\in f$, then only one path to $s_2$ is from $x_{s_1}$. But $px_{s_1}=h_1$ which is a contradiction.

If $x_{t_2}\notin f,x_{s_2}\in f$, then only one path to $t_2$ is from $x_{t_1}$. But $qx_{t_1}=h_2$ which is a contradiction.

If $x_{s_2}\notin f,x_{t_2}\notin f$, then only one path to $s_2,t_2$ is from $x_{s_1},x_{t_1}$. But $[\frac{gx_{s_1}x_{t_1}}{x_{s_2}x_{t_2}}]<[h_1]$ which is a contradiction.

$\bullet <(\boldsymbol{A},\boldsymbol{b}) \cdot f>_{\mathcal{O}_{h_2}\backslash \mathcal{O}_{h_1}\backslash \mathcal{D}} =0:$

(i) Case1: $h_2=gx_s$, $x_s\notin h_1$, set $f=q_1 q_2 q_3$, where $q_1\subset h_1\cap h_2$, $q_2\subset h_1\backslash h_2$, $q_3\subset h_2\backslash h_1$.

(a) $q_2=1$: $f=g\ or\ h_2$ which is a contradiction;

(b) $| ind(q_2)| \ge 2$: we choose $i\in ind(q_2)$ s.t. $i\ne max\{ind(h_1\backslash h_2)\}$, then $e_i=0$;

(c) $q_2=x_i$, $i= max\{ind(h_1\backslash h_2)\}$:

If $x_i$ is to empty in $g$, then $f\in \{gx_i,gx_ix_s\}$.

If $x_i$ is to some $x_j$ in $g$, then $f\in \{\frac{gx_i}{x_j},\frac{gx_ix_s}{x_j}\}$.

Because $x_s\in h_2\backslash h_1$, so $s<max\{ind(h_1\backslash h_2)\}=i$. And use $j<i$, we have $[f]<[h_2]$ which is a contradiction.

(ii) Case2: $h_2=px_{s_1}$, $g=px_{s_2}$, $s_2<s_1,x_{s_1}\notin h_1$, set $f=q_1 q_2 q_3$, where $q_1\subset h_1\cap h_2$, $q_2\subset h_1\backslash h_2$, $q_3\subset h_2\backslash h_1$.

(a) $q_2=1$: $f=h_2$ or $g$ which is a contradiction;

(b) $q_2=x_i$:

If $q_2=x_{s_2}$, then $f\in \{px_{s_2}$, $px_{s_1}x_{s_2}\}$ which is a contradiction or $f=\frac{px_{s_1}x_{s_2}}{x_j}$, $j\in ind(p)\cap (ind(h_2)\backslash ind(h_1))$ which satisfying $<(\boldsymbol{A},\boldsymbol{b})\cdot f>_g=a_{s_1 j}=0$;

If $q_2\ne x_{s_2}$, then $\left\{\begin{aligned}
& x_i\rightarrow \emptyset\ in\ g,then\ f\in \{gx_i,gx_ix_{s_1}\}.\\
& x_i\rightarrow x_{s_2},then\ f\in \{px_i,px_ix_{s_1}\}.\\
& x_i\rightarrow x_{j}\ in\ p,then\ f\in \{\frac{px_{s_2}x_i}{x_j},\frac{px_{s_2}x_ix_{s_1}}{x_j}\}.\\
\end{aligned}\right.$

When $i=max\{ind(h_1\backslash h_2)\}$: because $x_{s_1}\in h_2\backslash h_1$, so $s_1<max\{ind(h_1\backslash h_2)\}=i$. And use $j<i$, we have $[f]<[h_2]$ which is a contradiction;

When $i\ne max(h_1\backslash h_2)$, then $e_i=0$, so $<(\boldsymbol{A},\boldsymbol{b})\cdot f>_g=0$.

(c) $| ind(q_2)| =2$, $s_2 \in ind(q_2)$: let $i=max(h_1\backslash h_2)$,

then $\left\{\begin{aligned}
& x_i\rightarrow \emptyset\ in\ g,then\ f\in \{gx_i,gx_ix_{s_1}\}.\\
& x_i\rightarrow x_{j}\ in\ p,then\ f\in \{\frac{px_{s_2}x_i}{x_j},\frac{px_{s_2}x_ix_{s_1}}{x_j}\}.\\
\end{aligned}\right.$

(d) $| ind(q_2)| \ge 3$ or $| ind(q_2)| =2$, $s_2 \notin ind(q_2)$: we choose $i\in ind(q_2)$ s.t. $i\ne max(h_1\backslash h_2)$, then $e_i=0$.

$\bullet <(\boldsymbol{A},\boldsymbol{b}) \cdot f>_{\mathcal{O}_{h_1}\backslash \mathcal{O}_{h_2} \bigcap \mathcal{D}} =0$, $<(\boldsymbol{A},\boldsymbol{b}) \cdot f>_{\mathcal{O}_{h_1}\backslash \mathcal{O}_{h_2}\backslash \mathcal{D}} =0$ or a function of $e_{\mathcal{O}_{h_2}\backslash \mathcal{O}_{h_1}\backslash \mathcal{D}}$:

(i) Case1: $h_1=gx_s$

If $g\in \mathcal{O}_{h_1}\backslash \mathcal{O}_{h_2} \bigcap \mathcal{D}$ which means $x_s\in h_2$,

then $\left\{\begin{aligned}
\bullet & x_s\in f,e_s=0.\\
\bullet & x_s\notin f,all\ the\ paths\ to\ g\ only\ from\ h_2,\\
& but\ for\ i=max\{ind(h_1\backslash h_2)\},f\ can\ not\ generate\ it.\\
\end{aligned}\right.$

If $g\in \mathcal{O}_{h_1}\backslash \mathcal{O}_{h_2} \backslash \mathcal{D}$ which means $x_s\notin h_2$,

then $\left\{\begin{aligned}
\bullet & x_s\in f,g\ can\ not\ from\ x_s,set\ g\ is\ from\ q,\\
& so\ f=qx_s<gx_s=h_1\ which\ is\ a\ contradiction.\\
\bullet & x_s\notin f,all\ the\ paths\ to\ g\ only\ from\ h_2.\\
\end{aligned}\right.$

(ii) Case2: $h_1=px_{s_1}$, $g=px_{s_2}$, $s_2<s_1$

If $g\in \mathcal{O}_{h_1}\backslash \mathcal{O}_{h_2} \bigcap \mathcal{D}$ which means $x_{s_1}\in h_2$,

then $\left\{\begin{aligned}
\bullet & x_{s_1}\in f,e_{s_1}=0.\\
\bullet & x_{s_1}\notin f,all\ the\ paths\ to\ p\ only\ from\ h_2,\\
& but\ for\ i=max\{ind(h_1\backslash h_2)\},f\ can\ not\ generate\ it.\\
\end{aligned}\right.$

If $g\in \mathcal{O}_{h_1}\backslash \mathcal{O}_{h_2} \backslash \mathcal{D}$ which means $x_{s_1}\notin h_2$,

then $\left\{\begin{aligned}
\bullet & x_{s_1}\in f,set\ p\ is\ from\ q,so\ f=qx_{s_1}<px_{s_1}=h_1\\
& which\ is\ a\ contradiction.\\
\bullet & x_{s_1}\notin f,all\ the\ paths\ to\ g\ only\ from\ h_2.\\
\end{aligned}\right.$

\e

Now we let 

\begin{equation}
\mathcal{W}=\{g\in \mathcal{O}_{h_1}^c \cap \mathcal{O}_{h_2}^c\mid \exists\ p\in \mathcal{O}_{h_1}\backslash \mathcal{O}_{h_2}\backslash \mathcal{D}\ s.t.\ <(\boldsymbol{A},\boldsymbol{b}) \cdot g>_{p}\ne 0\}
\end{equation}

\section{the group $LTA(n,2)_{h_2}^{h_1}$ actions on $\mathcal{W}$}

\begin{lemma}\label{W unique}
For $(h_1, h_2)\in \Omega_{h_1,h_2}$, we have $|\mathcal{W}|\le 1$.

When $|\mathcal{W}|= 1$, we set $\mathcal{W}=\{g_\omega\}$, then 

\begin{equation}
\left\{
\begin{aligned}
&<(\boldsymbol{A},\boldsymbol{b}) \cdot h_1>_{g_\omega} =0 \\
&<(\boldsymbol{A},\boldsymbol{b}) \cdot h_2>_{g_\omega} =0\\
\end{aligned}\right.
\end{equation}
\end{lemma}

\pf

$\bullet |\mathcal{W}|\le 1$:

When $h_1=fx_s$, $x_s=max\{ind(h_1\backslash h_2)\}$:

(a) $| h_1\backslash h_2|=1$, $| h_2\backslash h_1|=2$: $g\in\mathcal{O}_{h_2}$ or $g=h_2$ which is a contradiction;

(b) $| h_1\backslash h_2|=2$, $| h_2\backslash h_1|=1$: $f\in\mathcal{O}_{h_1}\bigcap \mathcal{O}_{h_2}$ which is a contradiction;

(c) $| h_1\backslash h_2|=2$, $| h_2\backslash h_1|=2$: When $g$ have only one element in $h_2\backslash h_1$, $g\in\mathcal{O}_{h_2}$; When $g$ have all two elements in $h_2\backslash h_1$, $[g]<[h_2]$.

When $h_1=px_{s_1}$, $f=px_{s_2}$, $s_2<s_1$, $x_{s_1}=max(h_1\backslash h_2)$:

(a) $| h_1\backslash h_2|=1$, $| h_2\backslash h_1|=2$: When $g$ have only one element in $h_2\backslash h_1$, $g=\frac{h_2 x_{s_2}}{x_{j}}\ or\ \frac{h_2}{x_{j}}\in\mathcal{O}_{h_2}$, $x_{j}\in h_{2} \backslash h_1$; When $g$ have all two elements in $h_2\backslash h_1$, $[g]<[h_2]$;

(b) $| h_1\backslash h_2|=2$, $| h_2\backslash h_1|=1$: $[g]<[h_2]$ which is a contradiction;

(c) $| h_1\backslash h_2|=2$, $| h_2\backslash h_1|=2$: When $g$ have only one element in $h_2\backslash h_1$, $g=\frac{h_2 x_{h_1\backslash h_2\backslash s_1}}{x_{j}}\ or\ \frac{h_2 x_{s_2}}{x_{j}}\ or\ \frac{h_2}{x_{j}}\in\mathcal{O}_{h_2}$ or $g=\frac{h_2 x_{s_2} x_{h_1\backslash h_2\backslash s_1}}{x_{j}}$ which is the unique one, $x_{j}\in h_{2} \backslash h_1$; When $g$ have all two elements in $h_2\backslash h_1$, $[g]<[h_2]$.

$\bullet <(\boldsymbol{A},\boldsymbol{b}) \cdot h_1>_{g_\omega} =0$, $<(\boldsymbol{A},\boldsymbol{b}) \cdot h_2>_{g_\omega} =0$:

We know when $|\mathcal{W}|=1$, $| h_1\backslash h_2|=2$, $| h_2\backslash h_1|=2$ and $g_\omega=\frac{h_2 x_{s_2} x_{h_1\backslash h_2\backslash s_1}}{x_{j}}$.

so $deg(g)>deg(h_1)$, $deg(g)>deg(h_2)$ which means $<(\boldsymbol{A},\boldsymbol{b}) \cdot h_1>_{g_\omega} =0$ and $<(\boldsymbol{A},\boldsymbol{b}) \cdot h_2>_{g_\omega} =0$.

\e

\begin{lemma}\label{action on W}
With the same conditions of Theorem \ref{ltaour}, for any $(\boldsymbol{A},\boldsymbol{b})\in LTA(n,2)_{h_2}^{h_1}$, $\forall g\in \mathcal{W}$

\begin{equation}
\left\{
\begin{aligned}
&<(\boldsymbol{A},\boldsymbol{b}) \cdot f>_{g} =0,f\in \mathcal{O}^c_{h_1}\bigcap \mathcal{O}^c_{h_2}\backslash \mathcal{W} \\
&<(\boldsymbol{A},\boldsymbol{b}) \cdot f>_{g} =0,f\in \mathcal{O}_{h_2} \backslash \mathcal{O}_{h_1}\bigcap \mathcal{D} \\
\end{aligned}\right.
\end{equation}

\end{lemma}

\pf

$\bullet <(\boldsymbol{A},\boldsymbol{b}) \cdot f>_{g} =0$, $f\in \mathcal{O}^c_{h_1}\bigcap \mathcal{O}^c_{h_2}\backslash \mathcal{W}$:

Assume $<(\boldsymbol{A},\boldsymbol{b}) \cdot f>_{g} \ne 0$, then according to the definition of $\mathcal{W}$: $\exists$ $p\in \mathcal{O}_{h_1} \backslash \mathcal{O}_{h_2}\backslash \mathcal{D}$ s.t.
\begin{equation}
<(\boldsymbol{A},\boldsymbol{b}) \cdot g>_{p}\ne 0
\end{equation}

Then $<(\boldsymbol{A},\boldsymbol{b}) \cdot f>_{p}$ have a term $ <(\boldsymbol{A},\boldsymbol{b}) \cdot f>_{g}\cdot <(\boldsymbol{A},\boldsymbol{b}) \cdot g>_{p}\ne 0$, so $f\in \mathcal{W}$ which is a contradiction.

$\bullet <(\boldsymbol{A},\boldsymbol{b}) \cdot f>_{g} =0$, $f\in \mathcal{O}_{h_2} \backslash \mathcal{O}_{h_1}\bigcap \mathcal{D}:$

For $v\in \mathcal{O}_{h_1}\backslash \mathcal{O}_{h_2}\backslash \mathcal{D}$, $<(\boldsymbol{A},\boldsymbol{b}) \cdot g>_{v} \ne 0$,

(i) Case1: $h_2=fx_s,s\in ind(h_1)$, $h_1=vx_t,t\notin ind(h_2)$

Because $s\in v$, so any path from $i\in ind(f)$ to $x_s$, $a_{is}=0$ because $i,s \in h_2$. And $x_s\in g$ because $x_s\in h_1\cap h_2$ with no paths to itself.

(ii) Case2: $h_2=fx_s$, $s\in ind(h_1)$, $h_1=qx_{t_1}$, $v=qx_{t_2}$, $t_1\notin ind(h_2)$

Because $s\in q$: then any path from $i\in ind(v)$ to $x_s$, $a_{is}=0$ because $i,s \in h_2$. And $x_s\in g$ because $x_s\in h_1\cap h_2$ with no paths to itself.

(iii) Case3: $h_2=qx_{s_1}$, $f=qx_{s_2}$, $s_{2}<s_1$, $s_1\in ind(h_1)$, $h_1=vx_t$, $t\notin ind(h_2)$.

Because $s_1\in v$, then any path from $i\in ind(f)$ to $x_{s_1}$, $a_{is_1}=0$ because $i,s_1 \in h_2$. And $x_{s_1}\in g$ because $x_{s_1}\in h_1\cap h_2$ with no paths to itself.

(iv) Case4: $h_2=qx_{s_1}$, $f=qx_{s_2}$, $s_{2}<s_1$, $s_1\in ind(h_1)$, $h_1=px_{t_1}$, $v=px_{t_2}$, $t_2<t_1$, $t_1\notin ind(h_2)$.

Because $s_1\in p$: then any path from $i\in ind(f)$ to $x_{s_1}$,$a_{is_1}=0$ because $i=s_{2}\ or\ \in h_2$, $s_1 \in h_2$. And $x_{s_1}\in g$ because $x_{s_1}\in h_1\cap h_2$ with no paths to itself.
\e

\section{$LTA(n,2)_{h_2}^{h_1}$ is an automorphism group of $\mathbb{C}(\mathcal{B} \backslash \mathcal{W})$}
\begin{lemma}\label{coset-invariant}
With the same conditions of Theorem \ref{ltaour},the subcode $\mathbb{C}(\mathcal{B} \backslash \mathcal{W})$ is invariant under $LTA(n,2)_{h_2}^{h_1}$.
\end{lemma}

\pf

Because $\mathbb{C}(\mathcal{B} \backslash \mathcal{W})$ is generated by set $\mathcal{B} \backslash \mathcal{W}$, so we  only need to proof $\forall g\in \mathcal{B} \backslash \mathcal{W}$, $(\boldsymbol{A},\boldsymbol{b}) \cdot ev(g)\in \mathbb{C}(\mathcal{B} \backslash \mathcal{W})$ for $(\boldsymbol{A},\boldsymbol{b}) \in LTA(n,2)_{h_2}^{h_1}$.

Use another equation of the group action $LTA(n,2)$ \cite{alge}, we can write

\begin{equation}
(\boldsymbol{A},\boldsymbol{b})\cdot g=g+\mathop{\sum}_{q\in \mathcal{M}_n:q\prec g} u_q \cdot q
\end{equation}

And then from $C(\mathcal{I})$ as a decreasing monomial code, we get

\begin{equation}
(\boldsymbol{A},\boldsymbol{b})\cdot g=g+\mathop{\sum}_{q\in \mathcal{I}:q\prec g} u_q \cdot q
\end{equation}
 
Since 

\begin{equation}
\mathcal{B}=(\mathcal{O}_{h_2}\backslash \mathcal{O}_{h_1}\bigcap \mathcal{D})\bigcup (\mathcal{O}_{h_1}^c\bigcap \mathcal{O}_{h_2}^c)\bigcup \mathcal{T}
\end{equation}

\begin{equation}
\begin{aligned}
\mathcal{I}=& (\mathcal{O}_{h_1}\backslash \mathcal{O}_{h_2}\backslash \mathcal{D}) \bigcup (\mathcal{O}_{h_2}\backslash \mathcal{O}_{h_1}\backslash \mathcal{D})\\
& \bigcup (\mathcal{O}_{h_1}\bigcap \mathcal{O}_{h_2})\bigcup(\mathcal{O}_{h_1}\backslash \mathcal{O}_{h_2}\bigcap \mathcal{D}) \bigcup \mathcal{B}\\
\end{aligned}
\end{equation}

When $g\in \mathcal{T}$, this case is trivial;

When $g\in \mathcal{O}_{h_2}\backslash \mathcal{O}_{h_1}\bigcap \mathcal{D}$, $q\in \mathcal{I} \backslash \mathcal{B}$, according to Lemma \ref{action on Oh2}, we have $<(\boldsymbol{A},\boldsymbol{b})\cdot g>_{q}=0$;

When $g\in \mathcal{O}_{h_1}^c\bigcap \mathcal{O}_{h_2}^c\backslash \mathcal{W}$, $q\in \mathcal{I} \backslash \mathcal{B}$, according to Lemma \ref{action on Ohc1 Ohc2}, we have $<(\boldsymbol{A},\boldsymbol{b})\cdot g>_{q}=0$.

So we can write

\begin{equation}
(\boldsymbol{A},\boldsymbol{b})\cdot g=g+\mathop{\sum}_{q\in \mathcal{B} \backslash \mathcal{W}\prec g} u_q \cdot q.
\end{equation}

Therefore, $\forall g\in \mathcal{B} \backslash \mathcal{W}$, $(\boldsymbol{A},\boldsymbol{b}) \cdot ev(g)\in \mathbb{C}(\mathcal{B} \backslash \mathcal{W})$ for $(\boldsymbol{A},\boldsymbol{b}) \in LTA(n,2)_{h_2}^{h_1}$.\e

\section{Proofs of Theorem \ref{ltalemma}}\label{pf ltalemma}

Assume $\mathcal{W}^\prime=\{g_\omega\}$, and the proof is similar for $\mathcal{W}^\prime=\emptyset$.

Denote a initial coset by $S_0=ev(h_1)+ev(h_2)+ev(g_\omega)+\mathcal{C}(\mathcal{B}\backslash \mathcal{W})$.

We aim to proof the group action of $LTA(n,2)_{h_2}^{h_1}$ on $S_0$ can generate the whole set $\mathcal{S}_1$.

Let $(\boldsymbol{A},\boldsymbol{b}) \in LTA(n,2)_{h_2}^{h_1}$.

Utilizing Lemma \ref{action on h1} and Lemma \ref{W unique}, we obtain the equation of $(\boldsymbol{A},\boldsymbol{b})\cdot ev(h_1)$:

\begin{equation}\label{test1}
(\boldsymbol{A},\boldsymbol{b})\cdot ev(h_1)=ev(h_1)+\mathop{\sum}_{h\in \mathcal{O}_{h_1} \backslash \mathcal{O}_{h_2}\backslash \mathcal{D}} a_h \cdot ev(h)+\mathop{\sum}_{g\in \mathcal{B}\backslash \mathcal{W}} t_g \cdot ev(g)
\end{equation}

Utilizing Lemma \ref{action on h2} and Lemma \ref{W unique}, we obtain the equation of $(\boldsymbol{A},\boldsymbol{b})\cdot ev(h_2)$:

\begin{equation}\label{test2}
\begin{aligned}
(\boldsymbol{A},\boldsymbol{b})\cdot ev(h_2) & =ev(h_2)+\mathop{\sum}_{h\in \mathcal{O}_{h_2} \backslash \mathcal{O}_{h_1} \backslash \mathcal{D}} b_h \cdot ev(h)\\
& +\mathop{\sum}_{h\in \mathcal{O}_{h_1} \backslash \mathcal{O}_{h_2}\backslash \mathcal{D}} c_h \cdot ev(h)+\mathop{\sum}_{g\in \mathcal{B}\backslash \mathcal{W}} t_g \cdot ev(g)\\
\end{aligned}
\end{equation}

Utilizing Lemma \ref{action on Ohc1 Ohc2}, we obtain the equation of $(\boldsymbol{A},\boldsymbol{b})\cdot ev(g_\omega)$:

\begin{equation}\label{test3}
\begin{aligned}
(\boldsymbol{A},\boldsymbol{b})\cdot ev(g_\omega) & =ev(g_\omega)+\mathop{\sum}_{h\in \mathcal{O}_{h_1} \backslash \mathcal{O}_{h_2}\backslash \mathcal{D}} d_h \cdot ev(h)\\
& +\mathop{\sum}_{g_f\in \mathcal{B}\backslash \mathcal{W}} t_{g_f} \cdot ev(g_f)\\
\end{aligned}
\end{equation}

where $a_h\in e_{\mathcal{O}_{h_1} \backslash \mathcal{O}_{h_2}\backslash \mathcal{D}},b_h,c_h,d_h \in e_{\mathcal{O}_{h_2} \backslash \mathcal{O}_{h_1}\backslash \mathcal{D}}$.

Finally,

\begin{equation}\label{main1}
\begin{aligned}
(\boldsymbol{A},\boldsymbol{b})\cdot S_0 & =ev(h_1)+ev(h_2)+ev(g_\omega)+\mathop{\sum}_{h\in \mathcal{O}_{h_2}\backslash \mathcal{O}_{h_1} \backslash \mathcal{D}} b_h\cdot ev(h)\\
& +\mathop{\sum}_{h\in \mathcal{O}_{h_1}\backslash \mathcal{O}_{h_2} \backslash \mathcal{D}} (a_h+c_h+d_h)\cdot ev(h)\\
& +\mathop{\sum}_{g\in \mathcal{B}\backslash \mathcal{W}} a_g \cdot ev(g)+(\boldsymbol{A},\boldsymbol{b})\cdot \mathbb{C}(\mathcal{B}\backslash \mathcal{W})\\
\end{aligned}
\end{equation}

\begin{equation}\label{main2}
\begin{aligned}
& =ev(h_1)+ev(h_2)+ev(g_\omega)+\mathop{\sum}_{h\in \mathcal{O}_{h_2}\backslash \mathcal{O}_{h_1} \backslash \mathcal{D}} b_h\cdot ev(h)+\\
& \mathop{\sum}_{h\in \mathcal{O}_{h_1}\backslash \mathcal{O}_{h_2} \backslash \mathcal{D}} (a_h+c_h+d_h)\cdot ev(h)+\mathop{\sum}_{g\in \mathcal{B}\backslash \mathcal{W}} a_g \cdot ev(g)+\mathbb{C}(\mathcal{B}\backslash \mathcal{W})\\
\end{aligned}
\end{equation}

\begin{equation}\label{main3}
\begin{aligned}
& =ev(h_1)+ev(h_2)+ev(g_\omega)+\mathop{\sum}_{h\in \mathcal{O}_{h_2}\backslash \mathcal{O}_{h_1} \backslash \mathcal{D}} b_h\cdot ev(h)\\
& +\mathop{\sum}_{h\in \mathcal{O}_{h_1}\backslash \mathcal{O}_{h_2} \backslash \mathcal{D}} (a_h+c_h+d_h)\cdot ev(h)+\mathbb{C}(\mathcal{B}\backslash \mathcal{W})\\
\end{aligned}
\end{equation}

where

$\bullet$ (\ref{main1}) is from the three equations: (\ref{test1}), (\ref{test2}), (\ref{test3}); 

$\bullet$ (\ref{main2}) is from Lemma \ref{coset-invariant};

$\bullet$ (\ref{main3}) is because the absorption of $\mathbb{C}(\mathcal{B}\backslash \mathcal{W})$ about $\mathop{\sum}_{g\in \mathcal{B}\backslash \mathcal{W}} a_g \cdot ev(g)$.

Because $e_{\mathcal{O}_{h_1}\backslash \mathcal{O}_{h_2} \backslash \mathcal{D}}$ and $e_{\mathcal{O}_{h_2}\backslash \mathcal{O}_{h_1} \backslash \mathcal{D}}$ are independent, $(b_h,a_h+c_h+d_h)$ fills the entire space. In detail, we know when $e_{\mathcal{O}_{h_2}\backslash \mathcal{O}_{h_1} \backslash \mathcal{D}}$ are fixed which choose from $\{0,1\}^{\mid \mathcal{O}_{h_2}\backslash \mathcal{O}_{h_1} \backslash \mathcal{D} \mid }$, $b_{h},c_h,d_h$ are also determined. Then when $e_{\mathcal{O}_{h_1}\backslash \mathcal{O}_{h_2} \backslash \mathcal{D}}$ retrieve all values in $\{0,1\}^{\mid \mathcal{O}_{h_1}\backslash \mathcal{O}_{h_2} \backslash \mathcal{D} \mid }$, $\{a_h\}$ also retrieve all values in $\{0,1\}^{\mid \mathcal{O}_{h_1}\backslash \mathcal{O}_{h_2} \backslash \mathcal{D} \mid }$. And because $\{c_h\}$ and $\{d_h\}$ are fixed, so $\{a_h+c_h+d_h\}$ retrieve all values in $\{0,1\}^{\mid \mathcal{O}_{h_1}\backslash \mathcal{O}_{h_2} \backslash \mathcal{D} \mid }$.
		
Now we complete our proof for Theorem \ref{ltalemma}.\e		

\section{Proofs of Theorem \ref{ltaour}}\label{pf ltaour}

$\bullet$: When $\mathcal{W}=\emptyset$, this is equivalent to Theorem \ref{ltalemma};

$\bullet$:When $\mid \mathcal{W}\mid=1$:

$\forall S_1,S_2\in \mathcal{S}$:
\begin{equation}
S_1=S_1( \emptyset)+S_1( \mathcal{W})
\end{equation}

\begin{equation}
S_2=S_2( \emptyset)+S_2( \mathcal{W})
\end{equation}

Use Theorem \ref{ltalemma}, we have $S_1( \emptyset)$ and $S_2( \emptyset)$ have the same weight distribution, and $S_1( \mathcal{W})$ and $S_2(\mathcal{W})$ have the same weight distribution. And they are obtained through two different permutations. Combining $\mathcal{C}(\mathcal{B})=\{a\cdot g_\omega +\mathcal{C}(\mathcal{B}\backslash \{g_\omega\})|a\in \{0,1\}\}$, we get that $S_1$ and $S_2$ have the same weight distribution.
\e
\end{appendices}

\end{document}